# Simulations of the effects of tin composition gradients on the superconducting properties of Nb$_3$Sn conductors

L. D. Cooley[1,2], C. M. Fischer[1], P. J. Lee[1], and D. C. Larbalestier[1]

[1] *Applied Superconductivity Center, University of Wisconsin, Madison, Wisconsin 53706*
[2] *Materials Science Department, Brookhaven National Laboratory, Upton, New York, 11973*

In powder-in-tube (PIT) Nb$_3$Sn composites, the A15 phase forms between a central tin-rich core and a coaxial Nb tube, thus causing the tin content and superconducting properties to vary with radius across the A15 layer. Since this geometry is also ideal for magnetic characterization of the superconducting properties with the field parallel to the tube axis, a system of concentric shells with varying tin content was used to simulate the superconducting properties, the overall severity of the Sn composition gradient being defined by an index *N*. Using well-known scaling relationships and property trends developed in an earlier experimental study, the critical current density for each shell was calculated, and from this the magnetic moment of each shell was found. By summing these moments, experimentally measured properties such as pinning-force curves and Kramer plots could be simulated. We found that different tin profiles have only a minor effect on the shape of Kramer plots, but a pronounced effect on the irreversibility fields defined by the extrapolation of Kramer plots. In fact, these extrapolated values $H_K$ are very close to a weighted average of the superconducting properties across the layer for all *N*. The difference between $H_K$ and the upper critical field commonly seen in experiments is a direct consequence of the different ways measurements probe the simulated Sn gradients. Sn gradients were found to be significantly deleterious to the critical current density $J_c$, since reductions to both the elementary pinning force and the flux pinning scaling field $H_K$ compound the reduction in $J_c$. The simulations show that significant gains in $J_c$ of Nb$_3$Sn strands might be realized by circumventing strong compositional gradients of tin.

74.70.Ad, 74.25.Ha

## I. Introduction

Superconductivity in the A15 phase of the Nb-Sn system exists over the whole composition range from ~18 to ~25 at.% Sn, with better properties being found as stoichiometry is approached. Since all Nb$_3$Sn conductors are made by a diffusion reaction, the influence of not being at thermodynamic equilibrium should be considered, though in fact it seldom is. In reality it is almost certain that no filamentary Nb$_3$Sn composite of any design has ever been made to attain an equilibrium, that is uniform composition, because the need to maintain a small A15 grain size of 150 nm or less limits both the time and temperature of reaction. The consequence of restricting reaction, then, is that there is always a composition gradient across the superconducting layer. Devantay *et al.*[1] found that the critical temperature $T_c$ varies approximately linearly from about 6 to about 18 K. The variation of the upper critical field $H_{c2}$ with composition is less certain but also significant.[1,2,3] Since most modern high-field Nb$_3$Sn conductors introduce alloying elements to raise the resistivity and suppress the martensitic transformation[4], this variation may also be approximately linear with tin content too. Many years ago it was realized that compositional variations across the A15 phase would affect the observed superconducting properties[5,6,7,8,9,10,11]. However, neither the geometry of the largely bronze-route composites available in the 1980s, nor the analytical techniques of that time, made it feasible to survey in detail the actual composition gradients that were present across whole A15 layers.

Recently we returned to study this problem, incited by the availability of high quality powder-in-tube (PIT) Nb$_3$Sn composites made by Lindenhovius *et al.*[12,13] We find these PIT conductors to be of high quality both because of the very uniform nature of their filament tubes and because of their high critical current density $J_c$. These PIT conductors have two other important benefits for study, one being that the A15 phase forms layers that are 5-10 μm thick, suitable for accurate electron-probe microanalysis. The second advantage is a consequence of the special architecture of PIT wires, namely that the A15 layer forms by reaction between a central tin-rich core and a coaxial Nb tube, as shown in fig. 1. Thus the Sn content of the A15 layer is greater on the inside of the layer and superconductivity becomes weaker with radius outward from the core, making the system magnetically transparent in terms of superconducting shielding currents as a function of increasing field at given temperature. Magnetization measurements with the field coaxial to the strand are also relatively simple to interpret, because the superconducting A15 layer is a cylinder, which can be approximated by an array of coaxial tubes. Hawes *et al.*[14,15] used this geometry to estimate the radial variation of tin content in a binary PIT strand series by inverting the temperature dependence of the magnetic moment *m*, finding that the variation of $T_c$ with radius *r* was consistent with the radial variation of the tin content determined by energy-dispersive x-ray spectroscopy (EDX) in a scanning electron microsocpe.

In a subsequent study, Fischer *et al.*[16,17] studied the development of flux pinning and high-field superconducting properties by magnetization measurements in a vibrating sample magnetometer (VSM) at fields up to 14 T, which thus allowed a large range of magnetic field-temperature (*H-T*) space to be studied. A puzzling early finding was a large difference between the irreversibility field *H**, determined by either Kramer function extrapolation ($H_K$) or magnetization loop closure, and the upper critical field $H_{c2}$ for all heat treatment times. The difference between $H_{c2}$ and $H_K$ was always substantial but did decrease with increasing reaction, suggesting substantial changes in the tin composition gradient with increasing reaction. A curious feature of the Kramer plots, even those for very short reaction times when concentration gradients are largest, was that they were very linear all the way to zero bulk $J_c$ when measured at temperatures above ~12 K, at which the 14 T field of the VSM could completely suppress superconductivity in the wires.

This behavior is rather different from that of many bronze wires, which generally exhibit significantly curved Kramer plots. However, the normal architecture of bronze conductors places the highest Sn content at the *outside* of the filament, making the superconducting inhomogeneity magnetically invisible because the strongest shielding layer surrounds the weaker ones. Bronze conductors also have thin A15 layers (~1-2 µm) making accurate measurement of their Sn gradient difficult. The few such measurements[5,6,8,10,11] are all consistent with strong Sn gradients, making the association of a curved Kramer plot with strong Sn gradients across the whole layer plausible. Thus we were initially motivated to believe that the linear Kramer plots observed with PIT composites[16,18] resulted from rather flat Sn gradients. And in fact Hawes *et al.* showed that the Sn gradient in the PIT conductor was rather flat over most of the ~5 µm radius of the A15 layer, ranging from about 1 at.%Sn·µm$^{-1}$ at short reactions to as small as 0.4 at.%Sn.µm$^{-1}$ at longer times, the Sn content falling steeply to 18 at.% only near the Nb$_3$Sn/Nb interface.[14] Thus, while thermodynamics enforces the terminal compositions of the inner and outer radii of the A15 layer to be those in equilibrium with the Sn-rich phase and with Nb respectively, it does *not* dictate the nature of the gradient, which could vary from one architecture to another.

Based on EDX analyses, magnetic $T_c$ analyses and a volumetric specific heat $T_c$ analysis, Hawes *et al.* found that the Sn and $T_c$ gradients were both declining as the reaction proceeded. Fischer *et al.* and Godeke *et al.*[19] subsequently extended this zero-field study to make careful studies of how different in-field measurement techniques average the range of superconducting properties generated by the tin composition profile. The reasonable assumption made by Fischer *et al.* was that extrapolating Kramer plots from low fields defines an irreversibility field $H_K$ that is characteristic of a weighted average, rather than the best regions, of the A15 layer carrying the current. $H_{c2}$ can also be determined in the VSM by observing the change in the slope of the reversible magnetization as superconductivity is destroyed. However, inhomogeneity in the tin composition smears this transition such that the experimentally observed discontinuity at $H_{c2}$ represents the best region of the A15 layer. The difference between these two measurements is not trivial—at optimum reaction for highest $J_c$ and $H_K$, the actual measurements at 12 K showed that $H_K$ was 9.7 T while $H_{c2}$ was 13.1 T. Similarly large differences between $H_{c2}$ and *H** had been seen earlier by Suenaga *et al.*[20] in bronze conductors, although in that study the difference between $H_K$ and $H_{c2}$ was attributed to vortex melting effects in what was otherwise implied to be a homogeneous conductor.

To address the general influence of radial superconducting property gradients, we decided to simulate PIT strands using a system of concentric shells to represent various tin concentration profiles, drawn schematically in fig. 1. Current-density profiles associated with the input composition profiles were calculated based on literature values for $T_c$ and $H_{c2}$ and reasonable use of the Fietz and Webb scaling relations.[21] The magnetic moment of each shell could then be obtained from the current density and shell geometry and, by summing the contribution from all shells, we could estimate quantities that are obtained in magnetization experiments with the field parallel to the strand axis.

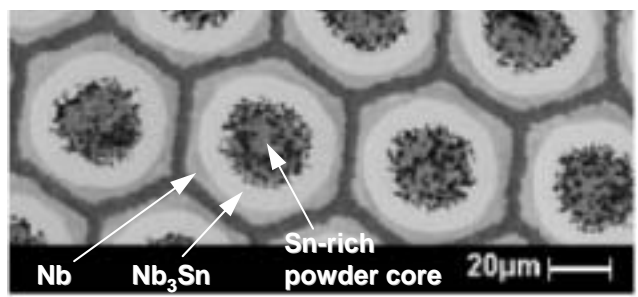

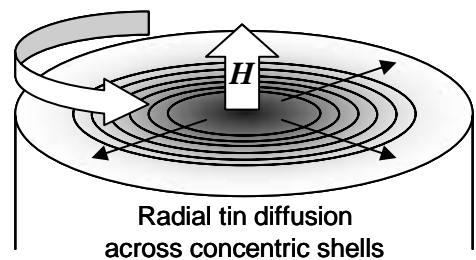

Fig. 1. Top: Scanning electron microscope image of a cross-section of a PIT Nb3Sn composite. Bottom: Schematic of a typical configuration for electromagnetic measurements, showing the arrangement of magnetization current shells with respect to the applied field and the radial direction of tin diffusion.

The general conclusion we draw from these simulations is that $H^*$, as determined by $H_K$, reflects a weighted *average*, while $H_{c2}$ reflects the *best* portion of the filaments, thus validating the assumption of Fischer *et al.* above. The difference between $H_{c2}$ and $H_K$ indeed diminishes as the Sn concentration gradient flattens. Broadly speaking, these simulations suggest that the *tin composition profile* determines $H^*$, suggesting that $H^*$ and $H_{c2}$ become *equal* in the absence of composition gradients. We here neglect thermal fluctuation effects. A very striking conclusion from the simulations is that the bulk flux pinning force is correlated with the properties of the weakest shells, even very far from $H_K$ or $H_{c2}$. This conclusion demonstrates that significant improvement of the $J_c$ of $Nb_3Sn$ will result from reducing or in the best of circumstances eliminating macroscopic tin composition gradients provided that the grain size can be maintained small at the same time.

## II. Experiment

A series of tin concentration profiles was generated using a spreadsheet. It was assumed that the inner radius was next to the tin source, anchoring the composition of the A15 layer at $Nb_{0.75}Sn_{0.25}$ at this interface, while the outer radius at the Nb/A15 interface was anchored at $Nb_{0.82}Sn_{0.18}$. Between these endpoints, the tin content was allowed to vary according to the expression

$$\%Sn\,(r) = 18 + 3.5\,[1 - r^N + (1-r)^N]. \quad (1)$$

The index $N$ indicates the severity and steepness of the overall gradient, where $N = 1$ represents a linear variation of tin composition and $N \to \infty$ describes a flat profile with an abrupt falloff near the $Nb_3Sn$ / Nb interface. A plot of the assumed tin composition profiles is shown in fig. 2.

The critical temperature at each shell was mapped according to each profile, assuming the linear dependence found by Devantay *et al.*[1]:

$$T_{c\text{-sim}}(r) = 6 + 12\,[(\%Sn(r) - 18)\,/\,7\,]\ \text{kelvin}. \quad (2)$$

These $T_{c\text{-sim}}$ profiles were used to generate profiles of upper critical field values $H_{c2\text{-sim}}(T,r)$ at $T = 0$ K,

$$\mu_0 H_{c2\text{-sim}}(0,r) = 0.69 \cdot 2.4 \cdot T_{c\text{-sim}}(r)\ \text{tesla}. \quad (3)$$

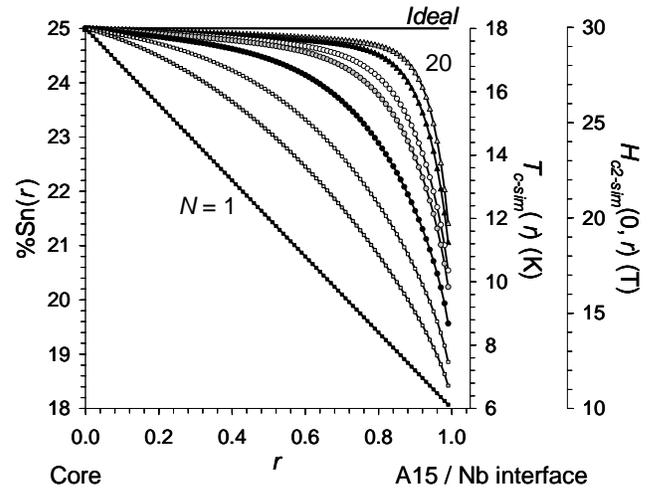

Fig. 2. Input profiles of tin composition as a function of radius. The index $N$ depicts the sharpness of the profile, where $N = 1, 2, 3, 5, 8, 10, 15,$ and $20$ are shown here and in figs. 4 to 7. An ideal profile, in which the tin content is constant at 25%, is also shown. Local values of the critical temperature and upper critical field (at $T = 0$) that correspond to the local A15 composition are indicated on the right axes.

Equation (3) is the usual dirty-limit expression $H_{c2}(0) = 0.69\,T_c\,(-dH_{c2}/dT|_{Tc})$ assuming the slope $\mu_0 dH_{c2}/dT$ at $T_c$ is $-2.4$ T/K. The particular value of the slope of $H_{c2}$ at $T_c$, being a bit higher than values in the literature,[22,23] was chosen here to make $H_{c2\text{-sim}}(0,0)$ close to 30 T, as suggested by recently observed values for unalloyed $Nb_3Sn$ prepared by hot isostatic pressing.[3] It should be noted that the variation of $H_{c2}$ with composition is not as well known as that of $T_c$. Finally, a temperature dependent factor of $1-(T/T_{c\text{-sim}})^{1.5}$ was multiplied by $H_{c2\text{-sim}}(0,r)$ used to get $H_{c2\text{-sim}}(T,r)$. This fit is an excellent approximation of the more complicated Maki-deGennes-Werthamer dirty-limit expression for $H_{c2}(T)$ [24,25]. In contrast to experiments, in which $T_c$ and $H_{c2}$ are single values determined by measurement, in the simulations the critical temperature and the upper critical field are *local* values; a hypothetical resistive or magnetization measurement would always produce experimental $T_c$ and $H_{c2}$ value that correspond to the maximum $T_{c\text{-sim}}$ value for all shells, *i.e.* 18 K and close to 30 T at 0 K for the shell next to the powder core.

As mentioned earlier, an interesting feature of the $J_c$ data on PIT composites is the observation of field and temperature scaling for a wide range of heat treatment conditions. Plots of $J_c^{1/2} H^{1/4}$ vs. field $H$ (Kramer plots) are linear for almost all heat treatment conditions[15] and measurement temperatures.[26] This permits the flux pinning force for each shell to be written

$$F_{p\text{-sim}}(h,T,r) = 3.5\,F_{max\text{-sim}}(T,r)\,h^{1/2}\,(1-h)^2 \quad (4)$$

where $h = H\,/\,H_{c2\text{-sim}}(T,r)$ and the factor 3.5 normalizes the field dependence. Further, analyses of the temperature dependence of the bulk pinning force showed that the temperature scaling could be expressed as $F_{max}(T) = 0.1 [H_K(T)]^{2.0}$, as shown in fig. 3. Therefore, the same temperature scaling for the bulk pinning force in each shell was assumed, namely:

$$F_{max-sim}(T, r) = 0.1 [\mu_0 H_{c2-sim}(T,r)]^2 \text{ GN/m}^3, \quad (5)$$

with $\mu_0 H_{c2-sim}$ in tesla. To obtain the critical current density for each shell, the definition of the Lorentz force was inverted,

$$J_{c-sim}(h,T,r) = F_{p-sim}(h,T,r) / h \text{ A/m}^2, \quad (6)$$

and $J_{c-sim}(h,T,r)$ was mapped to $J_{c-sim}(H,T,r)$ using $H = hH_{c2-sim}(T,r)$. At this point, it is important to distinguish that, while experimental data commonly exhibit scaling of the pinning force with $H^*$ or $H_K$, there is no *a priori* way to determine the scaling field from $H_{c2}$ in the absence of thermal fluctuations of the vortices. Moreover, flux-pinning models[27] suggest that the superconducting condensation energy, vortex density, and flux-lattice behavior should determine the pinning force, which are all tied to $H_{c2}$. Thus, eqs. 4 and 5 should be appropriate for simulating the local properties of each shell. This point will be discussed again later.

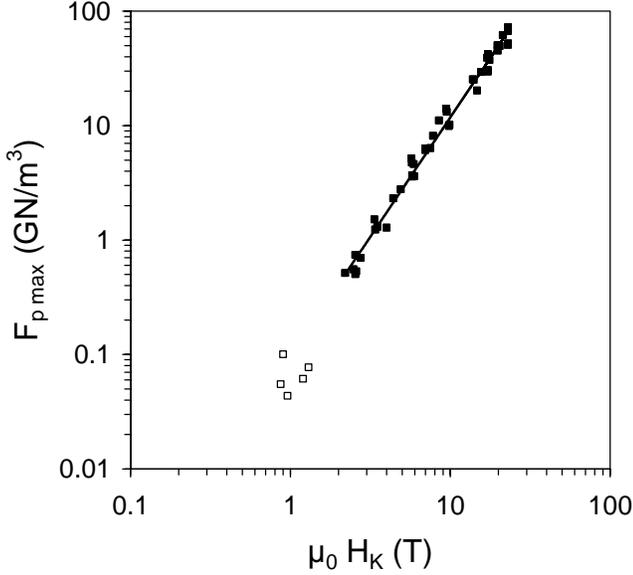

Fig. 3. Maximum bulk pinning force plotted against the irreversibility field, derived by extrapolating Kramer plots, for a series of 10 PIT samples, measured at 4.2 to 16 K. Despite differences in heat treatment duration (0.5 to 128 hours at 675°C ) and temperature, these data collapse onto the single linear fit given by Eq. 5. Data taken at 15 to 16 K (open symbols) yield very low values of $H_K$, but are still consistent with those at lower temperatures used for Eq. 5.

To get reasonable estimates for experimental quantities, the simulations were carried out by mapping the coordinate $r$ to a radius $R = R_{min} + 5r$ [μm]. This corresponds to a tin gradient that varies between a powder core located inside $R_{min} = 10$ μm and a Nb$_3$Sn/Nb interface located at $R_{max} = 15$ μm. These choices are typical radii observed by scanning electron microscopy (*e.g.* fig. 1). The magnetic moment $m_{sim}(H,T,R)$ for each shell was calculated from the critical current density by using the definition of a magnetic moment,

$$m_{sim}(H,T,R) = \pi R^2 I_{c-sim}(H,T,R) \quad (7)$$
$$= \pi R^2 t L J_{c-sim}(H,T,R) \text{ A·m}^2 \quad (8)$$

where $I_{c-sim}(H,T,R)$ is the current flowing around each shell, $t = (R_{max} - R_{min})/100$ is the thickness of each of the 100 shells. The total moment for all shells $m(H,T)$ was then obtained by summing the contributions from each shell.

Since the total magnetic moment is the physical property experimentally measured by magnetometry, the simulated profiles could be directly identified with experimental quantities. In particular, an estimate for the experimental critical current density $J_c$ could be obtained by applying the critical state model to the total moment and the input sample dimensions,

$$J_c(H,T) = 3 m(H,T) / [\pi L (R_{max}^3 - R_{min}^3)] \text{ A/m}^2. \quad (9)$$

This expression duplicates that used by Fischer *et al.* to determine the critical current density from vibrating sample magnetometer (VSM) measurements with the field parallel to the strand axis. Estimates for other experimental quantities, such as the bulk pinning force $F_p = \mu_0 H \cdot J_c$ and the Kramer function $F_K = J_c^{1/2} H^{1/4}$ were then derived.

Fig. 4 shows the simulated moment as a function of radius derived at 4.2 K, 12 T (plot a) and 12 K, 5 T (b), together with the mapping from $r$ to $R$. These temperatures correspond to temperatures investigated in the experiments of Fischer *et al.*, and the first plot also corresponds to a benchmark for recent R&D composite development.[28] Since a typical value for $L$ is a few millimeters, the simulated moment for each data point in fig. 4 corresponds to a real moment on the order of $10^{-10}$ A·m$^2$. The total moment for 100 shells in each PIT filament and ~200 filaments in a typical sample is thus about $2 \times 10^{-6}$ A·m$^2$ = $2 \times 10^{-3}$ emu, which is comparable to observed values for PIT composites in magnetometry experiments. The curve labeled "ideal" represents the so-far experimentally unattainable constant tin content of 25% across the entire A15 layer (*i.e.*, $N \rightarrow \infty$). Notice that there is no contribution to $m$ from many outer shells, even for very high $N$, when the chosen field and temperature exceed $T_{c-sim}$ and $H_{c2-sim}$.

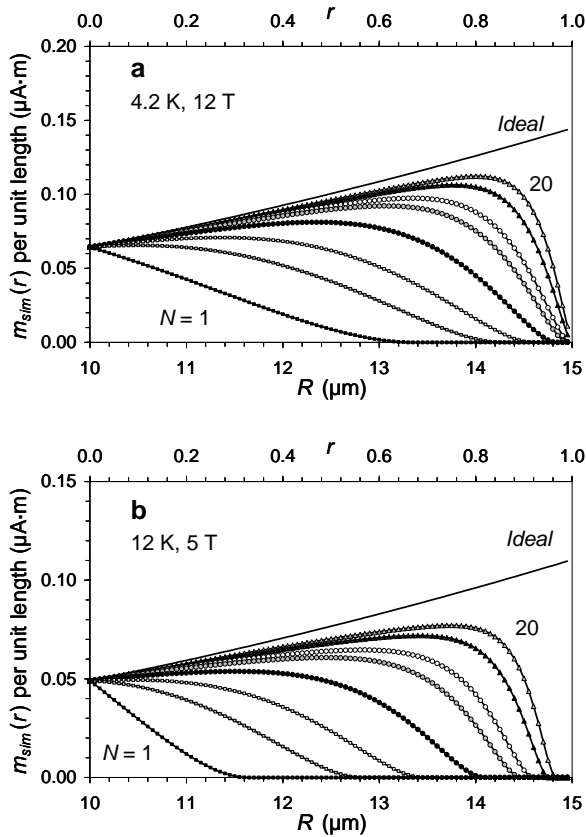

Fig. 4. Moment of the shells as a function of radius simulated at 4.2 K and 12 T (a), and at 12 K and 5 T (b). The indices for the curves are the same as for Fig. 2. Note that the coordinate $r$ has been mapped to a radius $R$ to simulate actual composites.

This is especially prominent at 12 K, where the outer sections of the layer are not even superconducting in zero field. The quantitative importance of the gradient to the moment accessible by experiments is very clear: If the tin concentration profile were indeed just a linear function of the thickness of the layer ($N = 1$), then the superconducting moment at 12 K would be small indeed due to the loss of many shells with $m_{sim} = 0$.

Figs. 5 through 7 present the experimental equivalent of $J_c(H)$, $F_p(H)$, and $F_K(H)$ for the same set of index values, with curves derived for 4.2 K and 12 K plotted in each figure. Consistent with Fig. 4, there is an evident loss of critical current density and bulk pinning force with decreasing $N$. The simulations also produce a shift of the peak of $F_p(H)$, and also in the field at which $F_p(H)$ goes to zero, towards higher fields as $N$ increases. After examining many such $m_{sim}(H,T,R)$ curves, we find that the curvature in the Kramer plots first appears for simulated fields that exceed $H_{c2-sim}(T,R_{max})$; that is, when the outermost shell is no longer superconducting. This is always true at 12 K, since $T_{c-sim}$ at $R_{max}$ is only 6 K for Nb18%Sn.

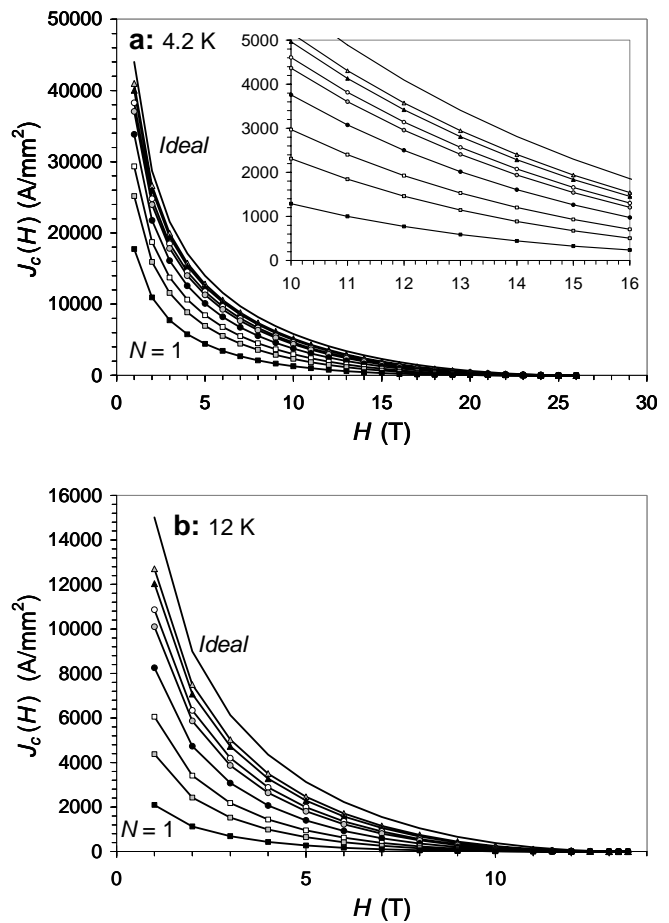

Fig. 5. Critical current density, as would be determined from a magnetization experiment of an actual PIT strand, as a function of field simulated at (a) 4.2 K, and (b) 12 K. The inset of plot (a) shows a magnified view of the mid field data. The indices for the curves follow the same sequence as in Fig. 2.

## III. Discussion

### A. Effect of tin gradients on the apparent irreversibility field

An important result of Fischer *et al.* was the finding of strong changes in, and large differences between, the values of $H_{c2}$ and $H_K$ as a function of heat treatment time.[16,17] The difference between $H_{c2}$ and $H_K$ at 12 K ranged from ~5 T at short times to ~3 T at long times, and both values increased with heat treatment time until saturating when the reaction was complete. It was argued that these differences were a direct consequence of the different ways in which $H_K$ and $H_{c2}$ are measured and tin composition profile, which was steep (low $N$) during the early stages of reaction and flattened (high $N$) as the reaction proceeded. The upper critical field was thought to represent the best (most tin-rich) regions of the A15 layer because $H_{c2}$ was defined by the return of the magnetization to the normal-state paramagnetic background at very high field, which occurs when the best potions of the A15 layer lose superconductivity. This was later found to be comparable to resistive measurements[24], which sample the best electrical pathway. By contrast, $H_K$, being derived from the full width of the hysteresis loop, was determined by the critical state across the entire A15 layer at lower fields, thus providing a weighted average of the superconducting properties across the layer. Since most Kramer plots obtained by Fischer *et al.* were highly linear, even close to the field axis, it was further assumed both that $H_K$ was well determined, and that the tin concentration profile was often rather flat except near the A15/Nb interface (equivalent to high $N$). However, it was not clear why this should be, especially for short heat treatment times. It is important to note that the measurement geometry used by Fischer *et al.* was such that currents enclose the powder core for $H < H_K$ and thus make large contributions to the measured magnetization, while for $H_K < H < H_{c2}$ hysteretic current loops might still be present, but their contribution to the overall magnetization, which includes reversible components, would hardly be detected. This makes confusion between these values in their experiment very unlikely.

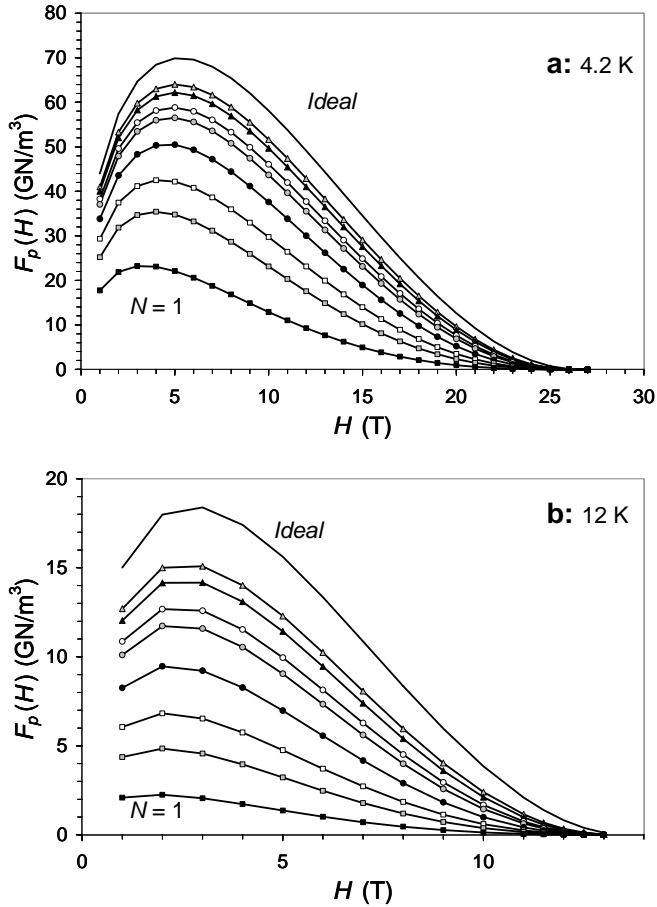

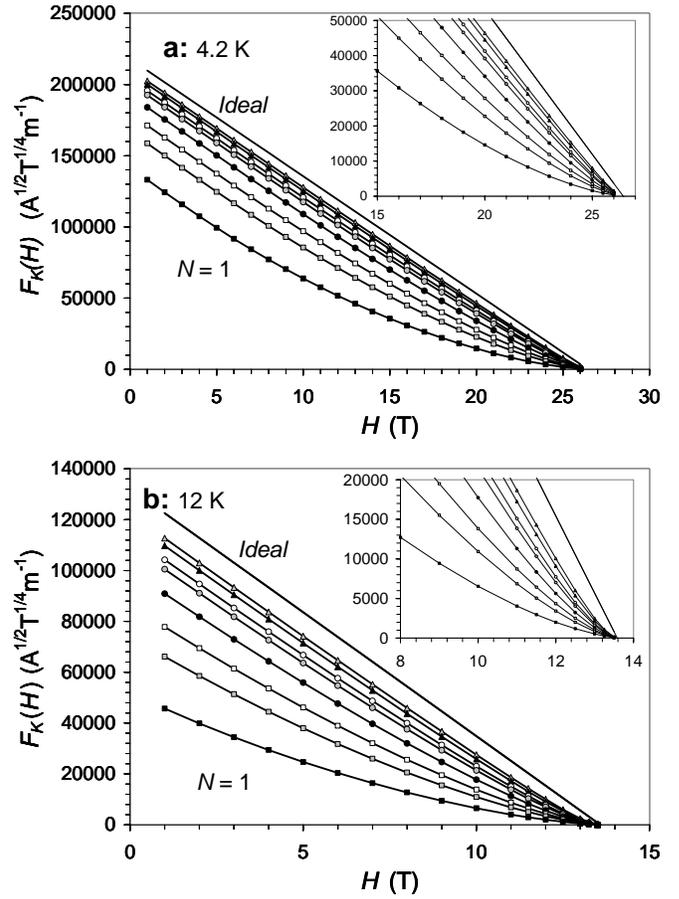

Fig. 6. Bulk flux-pinning force curves as a function of field simulated at (a) 4.2 K and (b) 12 K. The indices for the curves follow the same sequence as in Fig. 2.

Fig. 7. Kramer function curves as a function of field simulated at (a) 4.2 K and (b) 12 K. The indices for the curves follow the same sequence as in Fig. 2. The insets in both plots show a magnified view of the region near the field axis.

The Kramer plots derived from the simulation follow straight lines when all of the shells are superconducting. In many cases, especially when $N$ is high and $T$ is low, linear plots persist almost down to the field axis. Even when some shells are not superconducting because $H > H_{c2\text{-}sim}(T,r)$, as is often the case at high temperature, the curvature for large $N$ is not striking. On the other hand, the linear portions of the Kramer plots extrapolate to $H_K$ values less than the maximum value of $H_{c2\text{-}sim}(T)$ (being that for the inner shell with 25% Sn), even for high index values. This maximum value is clearly indicated on each plot by the curve labeled "ideal", since the ideal profile has no variation in tin composition and, by definition, produces $H_K = H_{c2}$ at all temperatures. Thus, the *primary influence of the different composition profiles is to alter the Kramer plot extrapolation*, and not to change greatly the Kramer plot shape itself. Moreover, the slopes of the linear portion of the Kramer plots at 12 K fall off with decreasing $N$, whereas they remain nearly constant at 4.2 K. This behavior is the same as that seen in the experimental curves of Fischer *et al.* The slope change is another effect of the loss of superconductivity in the outer shells far away from the tin source.

Table I. Comparison of the average value of the upper critical field with extrapolations of simulated Kramer functions, at 4.2 and 12 K.

| Gradient index $N$ | Weighted mean of $H_{c2\text{-}sim}(r)$ at 4.2 K (T) | $H_K$ at 4.2 K (T) | Weighted mean of $H_{c2\text{-}sim}(r)$ at 12 K (T) | $H_K$ at 12 K (T) |
|---|---|---|---|---|
| 1 | 15.7 | 19.1 | 7.1 | 9.6 |
| 2 | 19.4 | 21.1 | 8.5 | 10.3 |
| 3 | 21.2 | 22.1 | 9.4 | 10.8 |
| 5 | 23.0 | 23.3 | 10.5 | 11.4 |
| 8 | 24.2 | 24.2 | 11.3 | 11.9 |
| 10 | 24.6 | 24.6 | 11.7 | 12.1 |
| 15 | 25.2 | 25.1 | 12.2 | 12.5 |
| 20 | 25.5 | 25.4 | 12.6 | 12.7 |
| Ideal | 26.5 | 26.4 | 13.6 | 13.6 |

To test these predictions of Fischer *et al.* more explicitly, the $H_K$ values derived from the simulation are compared in Table I with the average of the $H_{c2\text{-}sim}(r)$ values across the simulated tin profiles at 4.2 and 12 K. The averaging is weighted so that only superconducting shells are considered; shells for which $T > T_{c\text{-}sim}(r)$ do not figure into the mean. Notice that there is very good agreement between $H_K$ and the weighted mean for all $N$ at 4.2 K. This is true even though the minimum value of $H_{c2\text{-}sim}(r)$ is only about 5 T at 4.2 K for $N = 1$, due to the low critical temperature of the tin-poor outer shells. At 12 K, the agreement is still reasonably good, and becomes better for increasing $N$. Thus, these simulations show that extrapolations of Kramer plots from low fields give good estimates of the *average* upper critical field across the active A15 layer, in agreement with the assertion by Fischer *et al.*.

In addition, at 4.2 K the difference between $H_{c2}$ ($= H_K$ for the ideal profiles) and $H_K$ is ~10 T for $N = 1$, and then rapidly drops to 3-4 T for intermediate $N$ and is only 1-2 T for $N \geq 10$. At 12 K, the separation is ~6 T for $N = 1$ and likewise falls to ~1 T for $N \geq 15$. This trend supports the conjecture by Fischer *et al.* that the tin composition profile flattens with increasing heat treatment time, which decreases the separation between $H_{c2}$ and $H_K$. This conclusion was also reached by Hawes *et al.*,[14,15] as we discuss in more detail later. Since the terminal separation of $H_{c2}$ and $H_K$ found by Fischer *et al.* was ~3 T at 12 K, the data in Table I suggest that a profile index of between 5 and 10 might apply to their optimum wires.

It is important to note that we did not simulate any effects of thermal fluctuations—the differences between $H_{c2}$ and $H_K$ are *entirely* due to the variation of $H_{c2\text{-}sim}(T,r)$ related to the simulated tin composition profile. Keeping in mind that different experiments probe local properties in different ways, this result suggests that the experimental difference between $H_{c2}$ and $H^*$ for real wires is mostly due to the effects of *tin composition inhomogeneity*. This implies that the $H^*(T)$ line is not determined by flux lattice properties in Nb$_3$Sn, as it is for high-temperature superconductors.

**B. Effects on flux pinning and the critical current density**

The tin composition profile can influence the global flux-pinning force $F_P$ by changing the magnitude of the local elementary pinning interaction and by shifting the scaling field for the bulk pinning-force curve.[27] The former effect is expressed in eq. (5), where the temperature dependence of the pinning force, derived from $H_{c2\text{-}sim}(T)$, reflects the condensation energy difference for a pinning interaction or the vortex line tension. The latter effect reflects flux density-dependent changes in summation and is expressed in eq. (4). Because these independent sources are compounded, the tin composition inhomogeneity actually has a significantly stronger effect on the flux pinning magnitude than on the critical field magnitudes, even though $J_c$ is generally evaluated far from $H_{c2}$. To illustrate this point, the maximum value of $F_p$ is summarized in Table II at 4.2 and 12 K. At 4.2 K, the maximum values are reduced from that of the ideal curve by ~28% for $N = 5$ and ~8% for $N = 20$ at 4.2 K. The corresponding decreases of $H_K$ are ~13% and ~4% for these respective indices in Table I, so the variations of the maximum value of $F_p$ are consistent with eq. 5. However, since $H_K$ is also lower for smaller $N$, the value of $F_p$ at *fixed* field suffers an additional loss. At 12 T, for example, $F_p$ and $J_c$ for $N = 5$ are ~40% less than their corresponding values for the ideal curve. The 12% additional subtraction is due to the shift in the scaling function (eq. 4) of each of t he simulated shells because of the corresponding lower $H_{c2\text{-}sim}(r)$ values.

It should be noted that, although flux-pinning theories express scaling functions in terms of $H_{c2}$, it is often useful to choose $H_K$ or $H^*$ as the scaling field (i.e. $h = H / H_K$ in eq. 4) in experiments. While this choice might seem to be the more rational one, properties of the flux-line lattice, from which various scaling relations are derived,[27] can be traced back directly to the Ginzburg-Landau theory,[29] suggesting $H_{c2}$ is the proper choice from a theoretical point of view. In the present simulation, these contradictory views are bridged because the *local* scaling function for each current shell, $F_{p\text{-}sim}(H,T,r)$ is tied to the corresponding local value of $H_{c2\text{-}sim}(T,r)$, whereas the *global* scaling curves $F_p(H,T)$ that emerge from the simulation (Fig. 6) appear to scale with $H/H_K$. This is an additional averaging effect of the current shells, where, as discussed above, $H_K$ represents the weighted mean of $H_{c2\text{-}sim}(r)$ values. Hence, *the averaging effect is why the irreversibility field is the appropriate scaling field for flux pinning*.

Table II. Variation of flux-pinning quantities with tin profile index.

| Gradient index $N$ | Maximum $F_p(H)$ at 4.2 K | | Maximum of $F_p(H)$ at 12 K | | $J_c$ at 12 T, 4.2 K | |
|---|---|---|---|---|---|---|
| | (GN/m$^3$) | % of ideal | (GN/m$^3$) | % of ideal | (A/mm$^2$) | % of ideal |
| 1 | 23 | 33 | 2.3 | 12 | 771 | 19 |
| 2 | 35 | 51 | 4.9 | 26 | 1460 | 36 |
| 3 | 42 | 61 | 6.8 | 37 | 1930 | 47 |
| 5 | 50 | 72 | 9.5 | 51 | 2500 | 61 |
| 8 | 56 | 81 | 12 | 64 | 2960 | 72 |
| 10 | 59 | 84 | 13 | 69 | 3140 | 77 |
| 15 | 62 | 89 | 14 | 77 | 3420 | 84 |
| 20 | 64 | 92 | 15 | 82 | 3570 | 87 |
| Ideal | 70 | -- | 18 | -- | 4090 | -- |

Because these simulations directly implicate the gradient of Sn in determining $J_c$, it is interesting to speculate how flattening the tin composition profile (while retaining similar grain size and grain-boundary density) could lead to gains in the performance of Nb$_3$Sn wires at 4.2 K and 12 T. Fischer et al. found that the highest critical current density of the A15 layer approached 3500 and 4300 A/mm$^2$ at 12 T, 4.2 K for a binary Nb$_3$Sn and a ternary (Nb,Ta)$_3$Sn PIT wire, respectively. When normalized to the entire tubular filament area, including the spent powder core and any unreacted Nb, these current densities are 1460 and 1780 A/mm$^2$ respectively. Since the simulation does not consider unreacted Nb but does include the core, the value of $J_{c\text{-}sim}$ should lie between these experimental values. As seen in Table II, this comparison suggests that the gradient index is perhaps 5 to 8, about the same range as that concluded from the critical field analyses discussed earlier. It should be kept in mind that the simulated values of $H_{c2}$ and $H_K$ (26.5 and >23 T) are close to the estimates of $H_K$(4.2 K) ~ 23 T for the binary composite and ~25 T for the ternary composite by Fischer et al.. While more experiments are needed to verify the actual values of $H_{c2}$ and $H_K$ at 4.2 K, this analysis suggests that modern PIT strands are operating at perhaps only 60% of the maximum $J_c$ attainable.

### C. Features of real tin composition profiles

More experimental focus has been given to electromagnetic characterization than to microchemical characterization of PIT strands, and particular information about the tin profiles in present state-of-the-art strands is lacking. Hawes et al.[14,15] conducted EDX analyses at various positions across the A15 layer for several strands from an earlier PIT conductor. The microprobe analyses showed that the composition fell off approximately linearly, from ~25% Sn near the powder core to ~22% near the A15/Nb interface. The final 4% drop to the tin-poor boundary of the A15 phase range could not be resolved by the instrument, suggesting that the moderate slope falls off more abruptly at the A15/Nb interface.

We simulated this profile by multiplying a factor of $(1 - cr)$ to Eq. (1) to approximate the observed linear profile, with $c = 0.5$ and $N = 20$. The simulated profile, shown in fig. 8a, is somewhat more severe than those suggested by the studies of Fischer et al. discussed earlier. Although the linear drop-off of tin content over much of the profile has a lower slope than the simulated profiles with $N < 5$ discussed earlier, the reduction of $J_c$ at 4.2 K shown in fig. 8b, is quite significant. At 12 T, $J_c$ reaches only 1600 A/mm$^2$, or about 39% of the ideal, no-gradient value of 4090 A/mm$^2$. The halving of the bulk pinning force is also evident in fig. 8c. However, the shape of the bulk pinning-force curve is almost identical to that of the ideal curve. Since the peak of the $h(1-h)^2$ function lies at $h = 0.2$, not only is the curvature consistent with the expected pinning function but the peak is also close to the expected position near 4/20. A slight tail in $F_p(H)$ can be seen above the irreversibility field at ~20 T, which is a common observation for real composites. The Kramer curve in fig. 8d is almost perfectly linear below ~14 T, the limit of many modern magnetometers or transport $J_c$ evaluations, but above this field $F_K(H)$ curves more strongly, resulting in a large difference between $H_K$ (20.5 T) and $H_{c2}$ (26.5 T). Taken together, all of these standard evaluations of flux pinning, if made in the laboratory at 4.2 K and below ~15 T, would suggest that this simulated Nb$_3$Sn example is from a strand of high quality. It is rather amazing that this occurs, even though the simulated results are, in fact, far from the ideal curves. Apparently, the strong contribution from magnetization currents flowing in the few stronger shells is very efficient at masking the deleterious contributions from the many weaker shells when all of the shells are averaged. This may be an artifact of the tubular geometry used in the simulation, and might not hold as strongly for the solid cylinders of most bronze or internal Sn conductors, perhaps resulting in more noticeable shifts in the $F_p(H)$ function and deviations from flux-pinning theories.

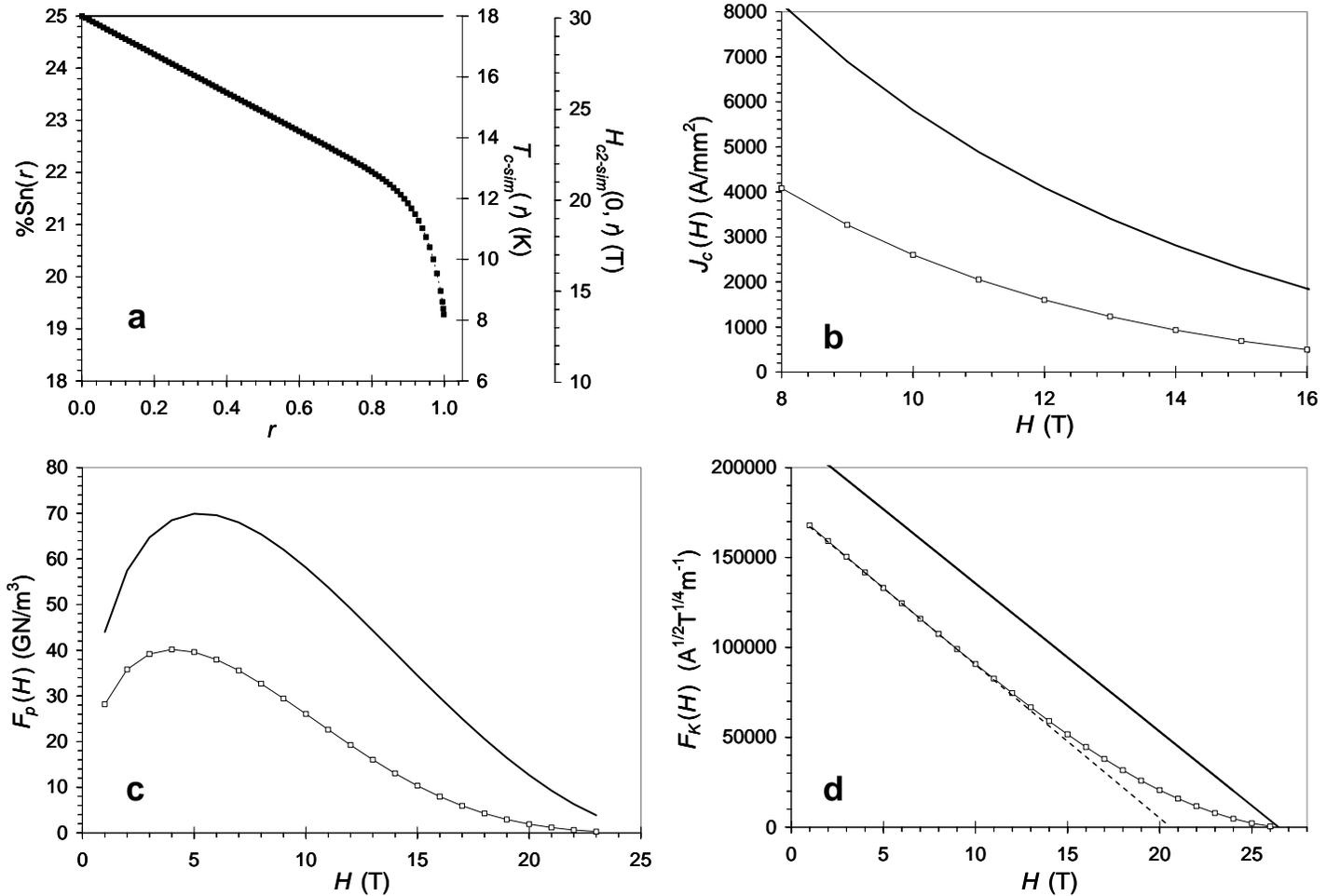

Fig. 8. Energy-dispersive x-ray spectroscopy analyses of a PIT strand in ref. 13 are simulated using the tin composition profile shown in plot (a), with corresponding local values of critical temperature and upper critical field indicated on the right axes. In plot (b), the simulated critical current density vs. field curve at 4.2 K is shown over a field range of technological interest. In plot (c), the corresponding bulk pinning force at 4.2 K is presented. Plot (d) shows the 4.2 K Kramer function, along with its extrapolation (dashed line) from data at 12 T and below. In all plots, the solid line represents the ideal profile of a constant 25% Sn.

The present simulations consider only radial tin gradients, whereas real composition profiles might contain circumferential and longitudinal gradients as well. These might arise due to disconnection of the powder core from the Nb tube, due to the formation of other intermetallic phases, or due to the preferential diffusion of tin along grain boundaries. In fact, one feature of experimental data that could not be reproduced by the simulation was the linearity of Kramer plots all the way down to the field axis at 12 K. We speculate that the cause may be due to circumferential or longitudinal Sn gradients in real wires. If the magnetization current is completely restricted to flow in concentric shells, circumferential variations in the tin composition would obstruct current and effectively switch off a given shell at a lower field than represented by its average composition. This would tend to amplify any effects of the radial compositional gradient. However, if the magnetization current is allowed to transfer between shells or migrate longitudinally, then the effect of circumferential obstructions would be blunted. Moreover, some shells might actually persist in carrying current at fields *above* that representative of their average composition. In the limit that all current shells have some circumferential variations, the effective network of current pathways could well remove the curvature from real Kramer plots altogether for sufficiently high gradient indices.

## IV. Conclusions

The variation of the tin composition in powder-in-tube $Nb_3Sn$ composites was modeled by analyzing a system of concentric shells. Different profiles of the local tin content were mapped, between the boundaries of Nb25%Sn at the A15/tin core interface and Nb18%Sn at the A15/Nb interface. The profile severity ranged from constant

Cooley *et al.*— submitted to *J. Appl. Phys.* 10

gradients ($N = 1$) to relatively flat plateaus with a steep gradient near the A15/Nb interface (large *N*).

By summing the calculated magnetic moment for each shell, simulations which yield experimentally accessible quantities were estimated. These simulations reproduced the experimental trends rather well, under the experimentally observed assumption that increasing heat treatment time for PIT strands enhances the value of the gradient index *N*. The experimental linear Kramer functions were reproduced well by the simulations for larger gradient indices. Extrapolations of the Kramer plots gave irreversibility fields $H_K$ that were very close to the weighted average of local upper critical field values corresponding to the input tin composition profiles. This suggests that Kramer plot extrapolations effectively *average* the superconducting properties across the A15 layer. An important result of the simulations is that this averaging property is passed on to the flux-pinning behavior, which explains why the global scaling field is $H_K$ even though flux pinning behavior on the elementary level is tied to the upper critical field $H_{c2}$. From analyses of Sn gradient-free, ideal profiles, for which Kramer plots extrapolate to $H_{c2}$, it was concluded that the difference between $H_{c2}$ and $H_K$ can be entirely attributed to the *tin composition profile*. The deleterious properties of severe tin gradients are compounded in quantities related to flux pinning, because both the elementary pinning force and the scaling field for summation are reduced. Thus, while $H_K$ is generally only 80% of $H_{c2}$ in modern Nb$_3$Sn composites, the critical current density at technologically important fields then reaches only ~60% of the available limit imposed by flux pinning for the stoichiometric A15 composition.

**Acknowledgments**

This work was supported by the US Department of Energy, Office of High Energy Physics. The facilities used for the work were partially supported by the NSF MRSEC for Nanostructured Materials and Interfaces at the University of Wisconsin.